\documentclass[12pt]{article}

\usepackage{amsmath,amssymb,amscd,amsbsy,amsgen,amsopn,amstext,
amsxtra}

\usepackage[dvips]{graphicx}

\begin{document}

\begin{flushright}
\end{flushright}

\begin{center}
{\Large{\bf Geometric Phase and Chiral Anomaly in Path Integral 
Formulation\footnote{Invited talk given at Path Integrals 2007, 
September 23-28, 2007, Max Planck Institute for Physics of Complex Systems, Dresden, Germany}}}
\end{center}
\vskip .5 truecm
\centerline{\bf  Kazuo Fujikawa }
\vskip .4 truecm
\centerline {\it Institute of Quantum Science, College of 
Science and Technology}
\centerline {\it Nihon University, Chiyoda-ku, Tokyo 101-8308, 
Japan}
\vskip 0.5 truecm


\begin{abstract}
All the geometric phases, adiabatic and non-adiabatic, are 
formulated in a unified manner in the second quantized path 
integral formulation. The exact hidden local symmetry inherent 
in the Schr\"{o}dinger equation defines the holonomy. All the 
geometric phases are shown to be topologically trivial. The 
geometric phases are briefly compared to the chiral anomaly 
which is naturally formulated in the path integral.
\end{abstract}


\section{ Second quantization}
To analyze various geometric phases in a unified 
manner~\cite{mead}-\cite{singh},
we start with an {\em arbitrary} complete basis set
\begin{eqnarray}
\int d^{3}x v_{n}^{\star}(t,\vec{x})v_{m}(t,\vec{x})=
\delta_{nm}\nonumber
\end{eqnarray}
and expand the field variable
as $\psi(t,\vec{x})=\sum_{n}b_{n}(t)v(t,\vec{x})$.
The action
\begin{eqnarray}
S=\int_{0}^{T}dtd^{3}x[
\psi^{\star}(t,\vec{x})i\hbar\frac{\partial}{\partial t}
\psi(t,\vec{x})-\psi^{\star}(t,\vec{x})
\hat{H}(\hat{\vec{p}}, \hat{\vec{x}},  
X(t))\psi(t,\vec{x})]
\end{eqnarray}
with background variables $X(t)=(X_{1}(t),X_{2}(t), ..)$ then 
becomes 
\begin{eqnarray}
S=\int_{0}^{T}dt\{\sum_{n}
b^{\star}_{n}(t)i\hbar\partial_{t}b_{n}(t)
-H_{eff} \}
\end{eqnarray}
with the effective Hamiltonian in the second quantized version
\begin{eqnarray}
\hat{H}_{eff}(t)&=&\sum_{n,m}\hat{b}_{n}^{\dagger}(t)[
\int d^{3}x v_{n}^{\star}(t,\vec{x})\hat{H}(\hat{\vec{p}},
\hat{\vec{x}}, X(t))v_{m}(t,\vec{x})
\nonumber\\
&&-\int d^{3}x v_{n}^{\star}(t,\vec{x})
i\hbar\frac{\partial}{\partial t}v_{m}(t,\vec{x})]
\hat{b}_{m}(t),
\end{eqnarray}
and $[\hat{b}_{n}(t), \hat{b}^{\dagger}_{m}(t)]_{\mp}
=\delta_{n,m}$, but statistics is not important in our 
application. We use fermions (Grassmann numbers) in the path 
integral.

The Schr\"{o}dinger picture $\hat{{\cal H}}_{eff}(t)$ is defined 
by replacing $\hat{b}_{n}(t)$ by 
$\hat{b}_{n}(0)$ in $\hat{H}_{eff}(t)$, and the 
evolution operator is given by~\cite{fuji-deguchi, fujikawa2}  
\begin{eqnarray}
\langle m|T^{\star}\exp\{-\frac{i}{\hbar}\int_{0}^{t}
\hat{{\cal H}}_{eff}(t)
dt\}|n\rangle
=\langle m(t)|T^{\star}\exp\{-\frac{i}{\hbar}\int_{0}^{t}
\hat{H}(\hat{\vec{p}}, \hat{\vec{x}},  
X(t))dt \}|n(0)\rangle\nonumber 
\end{eqnarray}
with time ordering symbol $T^{\star}$.
On the left-hand side 
$|n\rangle=\hat{b}_{n}^{\dagger}(0)|0\rangle$
and on the right-hand side
$\langle\vec{x}|n(t)\rangle=v_{n}(t,\vec{x})$.

The Schr\"{o}dinger probability amplitude with
$\psi_{n}(0,\vec{x})=v_{n}(0,\vec{x})$ is defined 
by~\cite{fujikawa2} 
\begin{eqnarray}
\psi_{n}(t,\vec{x})&=&\langle0|\hat{\psi}(t,\vec{x})
\hat{b}^{\dagger}_{n}(0)|0\rangle\nonumber\\
&=&\sum_{m} v_{m}(t,\vec{x})
\langle m|T^{\star}\exp\{-\frac{i}{\hbar}\int_{0}^{t}
\hat{{\cal H}}_{eff}(t)dt\}|n\rangle
\end{eqnarray}
and the path integral representation is given by
\begin{eqnarray}
&&\langle m|T^{\star}\exp\{-\frac{i}{\hbar}\int_{0}^{t}
\hat{{\cal H}}_{eff}(t)dt\}|n\rangle
=\int \prod_{n}{\cal D}b^{\star}_{n}{\cal D}b_{n}
\phi^{\star}_{m}(b^{\star}_{n}(t))\nonumber\\
&&\hspace{2cm}\times\exp\{\frac{i}{\hbar}\int_{0}^{t}dt[b^{\star}_{n}(t)
i\hbar\partial_{t}b_{n}(t)-H_{eff}(t)\}
\phi_{n}(b^{\star}_{n}(0))
\end{eqnarray}
with suitable  wave functions 
$\phi^{\star}_{m}(b^{\star}_{n}(t))$ and 
$\phi_{n}(b^{\star}_{n}(0))$ in the holomorphic 
representation~\cite{faddeev}. 
The general geometric terms automatically appear in the second
term of the 
{\em exact} $H_{eff}(t)$ (3) and thus the naive holomorphic 
wave functions are sufficient. This means that the analysis of 
geometric phases is reduced to a simple functional analysis in
the second quantized path integral. 

If one uses a specific basis 
$\hat{H}(\hat{\vec{p}}, \hat{\vec{x}},  
X(t))v_{n}(\vec{x};X(t))={\cal E}_{n}(X(t))v_{n}(\vec{x};X(t))$
and assumes "diagonal dominance" in the effective Hamiltonian in
(5), we have the adiabatic formula
\begin{eqnarray}
\psi_{n}(t,\vec{x})
\simeq v_{n}(\vec{x};X(t))
\exp\{-\frac{i}{\hbar}\int_{0}^{t}[{\cal E}_{n}(X(t))
-v_{n}^{\star}i\hbar\frac{\partial}{\partial t}v_{n}dt\}
\end{eqnarray}
which shows that the {\em adiabatic approximation} is 
equivalent to the {\em approximate diagonalization} of 
$H_{eff}$, and thus the geometric phases are 
{\em dynamical}~\cite{fuji-deguchi, fujikawa2}.

\section{Hidden local gauge symmetry}

Since $\psi(t,\vec{x})=\sum_{n}b_{n}(t)v_{n}(t,\vec{x})$, we 
have an exact local symmetry~\cite{fujikawa2}
\begin{eqnarray}
&&v_{n}(t,\vec{x})\rightarrow v^{\prime}_{n}(t, \vec{x})=
e^{i\alpha_{n}(t)}v_{n}(t,\vec{x}),\nonumber\\
&&b_{n}(t) \rightarrow b^{\prime}_{n}(t)=
e^{-i\alpha_{n}(t)}b_{n}(t), \ \ \ \ n=1,2,3,... .
\end{eqnarray}
This symmetry means an arbitrariness in the choice of the 
coordinates in the functional space.
The exact Schr\"{o}dinger amplitude $\psi_{n}(t,\vec{x})
=\langle0|\hat{\psi}(t,\vec{x})\hat{b}^{\dagger}_{n}(0)
|0\rangle$ is transformed under this substitution rule as 
\begin{eqnarray} 
\psi^{\prime}_{n}(t,\vec{x})=e^{i\alpha_{n}(0)}
\psi_{n}(t,\vec{x})\nonumber
\end{eqnarray}
for any $t$. Namely, we have the ray representation with a 
constant phase change, and we have 
an enormous hitherto unrecognized exact hidden symmetry behind 
the ray representation.

The combination $\psi_{n}(0,\vec{x})^{\star}\psi_{n}(T,\vec{x})$
 thus becomes manifestly gauge invariant. For example, for 
adiabatic approximation (6) we have
\begin{eqnarray}
\psi_{n}(0,\vec{x})^{\star}\psi_{n}(T,\vec{x})
&=&v_{n}(0,\vec{x}; X(0))^{\star}v_{n}(T,\vec{x};X(T))
\nonumber\\
&&\times\exp\{-\frac{i}{\hbar}\int_{0}^{T}[{\cal E}_{n}(X(t))
-v^{\star}_{n}i\hbar\frac{\partial}{\partial t}v_{n}]dt\}
\end{eqnarray}
If one chooses a specific gauge 
$v_{n}(T,\vec{x};X(T))=v_{n}(0,\vec{x}; X(0))$, the prefactor 
$v_{n}(0,\vec{x}; X(0))^{\star}v_{n}(T,\vec{x};X(T))$ is real 
and 
positive. The factor in the exponential then represents the 
entire gauge invariant phase. 
\\

\noindent {\bf Parallel transport and holonomy}\\
\\
The parallel transport of $v_{n}(t,\vec{x})$ is defined by
\begin{eqnarray}
\int d^{3}x v^{\dagger}_{n}(t,\vec{x})\frac{\partial}{\partial t}
v_{n}(t,\vec{x})=0\nonumber
\end{eqnarray}
which follows from 
$\int d^{3}x v^{\dagger}_{n}(t,\vec{x})
v_{n}(t+\delta t,\vec{x})={\rm real\ and\ positive}$, 
and \\
$\int d^{3}x v^{\dagger}_{n}(t+\delta t,\vec{x})
v_{n}(t+\delta t,\vec{x})
=\int d^{3}x v^{\dagger}_{n}(t,\vec{x})
v_{n}(t,\vec{x})$.
By using the hidden local gauge
$\bar{v}_{n}(t,\vec{x})=e^{i\alpha_{n}(t)}v_{n}(t,\vec{x})$
for general $v_{n}(t,\vec{x})$, one may impose the condition
\begin{eqnarray}
\int d^{3}x \bar{v}^{\dagger}_{n}(t,\vec{x})\frac{\partial}
{\partial t}
\bar{v}_{n}(t,\vec{x})=0
\end{eqnarray}
which gives
\begin{eqnarray}
\bar{v}_{n}(t,\vec{x})
=\exp[i\int_{0}^{t}dt^{\prime}
\int d^{3}x v^{\dagger}_{n}(t^{\prime},\vec{x})
i\partial_{t^{\prime}}v_{n}(t^{\prime},\vec{x})]v_{n}(t,\vec{x})
.
\end{eqnarray}
The {\em holonomy} for a cyclic motion  is then defined by  
\begin{eqnarray}
&&\bar{v}^{\dagger}_{n}(0,\vec{x})
\bar{v}_{n}(T,\vec{x})\nonumber\\
&&=v^{\dagger}_{n}(0,\vec{x})v_{n}(T,\vec{x})
\exp[i\int_{0}^{T}dt^{\prime}
\int d^{3}x v^{\dagger}_{n}(t^{\prime},\vec{x})
i\partial_{t^{\prime}}v_{n}(t^{\prime},\vec{x})].
\end{eqnarray}
This holonomy of {\em basis vectors}, not Schr\"{o}dinger 
amplitude, determines {\em all} the geometric phases in the 
second quantized formulation~\cite{fujikawa4}. 

\section{ Non-adiabatic phase}

\noindent{\bf (i) Cyclic evolution}:\\
The cyclic evolution is defined by 
$\psi(T,\vec{x})=e^{i\phi}\psi(0,\vec{x})$ or by 
\begin{eqnarray} 
\psi(t,\vec{x})=e^{i\phi(t)}\tilde{\psi}(t,\vec{x}),\ \ \
\tilde{\psi}(T,\vec{x})=\tilde{\psi}(0,\vec{x})
\end{eqnarray}
with $\phi(T)=\phi,
 \ \ \phi(0)=0$.
If one chooses the first element of the arbitrary basis set 
$\{v_{n}(t,\vec{x})\}$ such that
$v_{1}(t,\vec{x})=\tilde{\psi}(t,\vec{x})$, one has diagonal
$\hat{H}_{eff}(t)$ and
\begin{eqnarray}
\psi(t,\vec{x})
&=&v_{1}(t,\vec{x})\exp\{-\frac{i}{\hbar}
[\int_{0}^{t}dt\int d^{3}x v^{\star}_{1}(t,\vec{x})
\hat{H}v_{1}(t,\vec{x})\nonumber\\
&&\hspace{1.5 cm}-\int_{0}^{t}dt\int d^{3}x 
v^{\star}_{1}(t,\vec{x})i\hbar\partial_{t}v_{1}(t,\vec{x})]\}
\end{eqnarray}
in (4).
Under the hidden local symmetry, we have
$\psi(t,\vec{x})\rightarrow e^{i\alpha_{1}(0)}\psi(t,\vec{x})$
and the gauge invariant quantity
\begin{eqnarray}
\psi^{\dagger}(0,\vec{x})\psi(T,\vec{x})
&=&v^{\star}_{1}(0,\vec{x})v_{1}(T,\vec{x})
\exp\{-\frac{i}{\hbar}
\int_{0}^{T}dt\int d^{3}x[ v^{\star}_{1}(t,\vec{x})
\hat{H}v_{1}(t,\vec{x}) \nonumber\\
&&\hspace{5 cm} -v^{\star}_{1}(t,\vec{x})i\hbar\partial_{t}
v_{1}(t,\vec{x})]\}
\end{eqnarray}
If one chooses $v_{1}(0,\vec{x})=v_{1}(T,\vec{x})$,
$v^{\star}_{1}(0,\vec{x})v_{1}(T,\vec{x})$ becomes real and 
positive, and the factor~\cite{fujikawa3}
\begin{eqnarray}
\beta=\oint dt \int d^{3}x v^{\star}_{1}(t,\vec{x})
i\frac{\partial}{\partial t}v_{1}(t,\vec{x})
\end{eqnarray}
gives the {\em non-adiabatic phase}\cite{aharonov}.

Note that the so-called "projective Hilbert space" and the transformation
\begin{eqnarray}
\psi(t,\vec{x})\rightarrow e^{i\omega(t)}\psi(t,\vec{x})
,\nonumber
\end{eqnarray}
which is  not the symmetry of the Schr\"{o}dinger equation, is 
not used in our formulation. 
Also, the holonomy of the basis vector, not the Schr\"{o}dinger
amplitude,  determines the non-adiabatic phase.
Our derivation of the non-adiabatic phase (15), which 
works in the path integral (5) also, is quite 
different from that in~\cite{aharonov}.  
\\

\noindent{\bf (ii) Non-cyclic evolution}:\\  
It is shown that {\em any} exact solution of the Schr\"{o}dinger 
equation is written in the form~\cite{fujikawa4},
\begin{eqnarray}
\psi_{k}(\vec{x},t)
&=&v_{k}(\vec{x},t)\exp\{-\frac{i}{\hbar}\int_{0}^{t}
\int d^{3}x[v^{\dagger}_{k}(\vec{x},t)
\hat{H}(t)v_{k}(\vec{x},t)\nonumber\\
&&\hspace{3 cm}-v^{\dagger}_{k}(\vec{x},t)
i\hbar\frac{\partial}{\partial t}v_{k}(\vec{x},t)]\}
\end{eqnarray}
if one suitably chooses the basis set $\{v_{k}(\vec{x},t)\}$,
though the periodicity is generally lost, $v_{k}(\vec{x},0)
\neq v_{k}(\vec{x},T)$.
$\psi_{k}(t,\vec{x})$ is transformed as
$\psi_{k}(t,\vec{x})\rightarrow e^{i\alpha_{k}(0)}
\psi_{k}(t,\vec{x})$ under the hidden local symmetry,
and 
\begin{eqnarray}
&&\int d^{3}x
\psi_{k}^{\dagger}(0,\vec{x})\psi_{k}(T,\vec{x})
=\int d^{3}x
v_{k}^{\dagger}(0,\vec{x})v_{k}(T,\vec{x})\nonumber\\
&&\times\exp\{\frac{-i}{\hbar}\int_{0}^{T}dtd^{3}x[
v_{k}^{\dagger}(t,\vec{x})\hat{H}(t)v_{k}(t,\vec{x})
-v_{k}^{\dagger}(t,\vec{x})i\hbar\partial_{t}
v_{k}(t,\vec{x})]\}
\end{eqnarray}
is thus manifestly gauge invariant. 
By choosing a suitable hidden symmetry 
$v_{k}(t,\vec{x})\rightarrow
e^{i\alpha_{k}(t)}v_{k}(t,\vec{x})$, one can make $\int d^{3}x
v_{k}^{\dagger}(0,\vec{x})v_{k}(T,\vec{x})$ real and positive.
Then the exponential factor defines the non-cyclic 
non-adiabatic phase\cite{samuel}. The present definition also
works in the path integral (5).

It is shown that geometric phases for mixed 
states~\cite{sjoqvist, singh} are similarly
 formulated in the second quantized formulation~\cite{fujikawa4}.

\section{ Exactly solvable example}
We study the model 
\begin{eqnarray}
\hat{H}=-\mu\hbar\vec{B}(t)\vec{\sigma},\ \ 
\vec{B}(t)=B(\sin\theta\cos\varphi(t), 
\sin\theta\sin\varphi(t),\cos\theta )
\end{eqnarray}
with $\varphi(t)=\omega t$ and constant $\omega,\ B$ and  
$\theta$.

To {\em diagonalize} the second quantized $H_{eff}$, we define
the constant $\alpha$ by~\cite{fujikawa4}  
\begin{eqnarray}
\tan\alpha=\frac{\hbar\omega\sin\theta}{2\mu\hbar B+\hbar\omega
\cos\theta}
\end{eqnarray}
or equivalently
$2\mu\hbar B\sin\alpha=\hbar\omega\sin(\theta-\alpha)$, and the 
basis vectors
\begin{eqnarray}
w_{+}(t)=\left(\begin{array}{c}
            \cos\frac{1}{2}(\theta-\alpha) e^{-i\varphi(t)}\\
            \sin\frac{1}{2}(\theta-\alpha)
            \end{array}\right),\  
w_{-}(t)=\left(\begin{array}{c}
            \sin\frac{1}{2}(\theta-\alpha) e^{-i\varphi(t)}\\
            -\cos\frac{1}{2}(\theta-\alpha)
            \end{array}\right)\nonumber
\end{eqnarray}
which satisfy $w_{\pm}(0)=w_{\pm}(T)$ with 
$T=\frac{2\pi}{\omega}$, and
\begin{eqnarray}
w_{\pm}^{\dagger}(t)\hat{H}w_{\pm}(t)
=\mp \mu\hbar B\cos\alpha,\ \
w_{\pm}^{\dagger}(t)i\hbar\partial_{t}w_{\pm}(t)
=\frac{\hbar\omega}{2}(1\pm\cos(\theta-\alpha))
\nonumber
\end{eqnarray}
The effective Hamiltonian $H_{eff}$ is now diagonal, and 
the {\em exact} solution of the Schr\"{o}dinger equation 
$i\hbar\partial_{t}\psi(t)=\hat{H}\psi(t)$ is given 
by~\cite{fujikawa4}
\begin{eqnarray}
\psi_{\pm}(t)
=w_{\pm}(t)\exp\{-\frac{i}{\hbar}\int_{0}^{t}dt^{\prime}
[w_{\pm}^{\dagger}(t^{\prime})\hat{H}w_{\pm}(t^{\prime})
-w_{\pm}^{\dagger}(t^{\prime})i\hbar\partial_{t^{\prime}}
w_{\pm}(t^{\prime})]\}.
\end{eqnarray}
This exact solution may be regarded either as an exact version
of the adiabatic phase or as the cyclic nonadiabatic phase 
in our formulation.

We examine some limiting cases of this exact solution:\\ 
\noindent (i) For {\em adiabatic limit} 
$\hbar\omega/(\hbar\mu B)\ll 1$, we have from (19)
\begin{eqnarray}
\alpha\simeq[\hbar\omega/2\hbar\mu B]\sin\theta, \nonumber
\end{eqnarray}
and if one sets 
$\alpha=0$ in (20), one recovers the Berry's phase~\cite{berry}
\begin{eqnarray}
\psi_{\pm}(T)\simeq\exp\{i\pi(1\pm\cos\theta) \}
\exp\{\pm\frac{i}{\hbar}\int_{0}^{T}dt
\mu\hbar B\}w_{\pm}(T)
\end{eqnarray}
with
\begin{eqnarray}
w_{+}(t)=\left(\begin{array}{c}
            \cos\frac{1}{2}\theta e^{-i\varphi(t)}\\
            \sin\frac{1}{2}\theta
            \end{array}\right), \ \  
w_{-}(t)=\left(\begin{array}{c}
            \sin\frac{1}{2}\theta e^{-i\varphi(t)}\\
            -\cos\frac{1}{2}\theta
            \end{array}\right).\nonumber
\end{eqnarray}

\noindent (ii) For {\em non-adiabatic} limit 
$\hbar\mu B/(\hbar\omega)\ll 1$, we have from (19)
\begin{eqnarray}
\theta-\alpha\simeq[2\hbar\mu B/\hbar\omega]\sin\theta
\nonumber
\end{eqnarray}
and if one sets $\alpha=\theta$ in (20), one obtains the 
trivial geometric phase
\begin{eqnarray}
\psi_{\pm}(T)
&\simeq&w_{\pm}(T)\exp\{\pm\frac{i}{\hbar}\int_{0}^{T}dt
[\mu\hbar B\cos\theta]\}
\end{eqnarray}
with
\begin{eqnarray}
w_{+}(t)=\left(\begin{array}{c}
            e^{-i\varphi(t)}\\
            0
            \end{array}\right), \ \ 
w_{-}(t)=\left(\begin{array}{c}
            0\\
            -1
            \end{array}\right).\nonumber
\end{eqnarray}
This shows that the ``monopole-like'' phase in (21) is smoothly 
connected to a trivial phase in the exact solution, and thus 
the geometric phase is 
{\em topologically trivial}~\cite{fuji-deguchi}.

\section{ Chiral anomaly}

It is known that all the anomalies in gauge field 
theory~\cite{bell,adler} are understood in the
path integral as arising from the non-trivial Jacobians under 
symmetry transformations~\cite{fujikawa,fujikawa-suzuki}. For 
example, in the fermionic path integral
\begin{eqnarray}
\int {\cal D}\bar{\psi}{\cal D}\psi \exp\{i\int d^{4}x
[\bar{\psi}i\gamma^{\mu}(\partial_{\mu}-igA_{\mu})\psi]\}
\end{eqnarray}
and for infinitesimal chiral transformation 
\begin{eqnarray}
\psi(x)\rightarrow e^{i\omega(x)\gamma_{5}}\psi(x),\ \ \ \ 
\bar{\psi}(x)\rightarrow \bar{\psi}(x)e^{i\omega(x)\gamma_{5}},
\nonumber
\end{eqnarray}
we have
\begin{eqnarray}
{\cal D}\bar{\psi}{\cal D}\psi\rightarrow 
\exp\{-i\int d^{4}x \omega(x)\frac{g^{2}}{16\pi^{2}}
\epsilon^{\mu\nu\alpha\beta}F_{\mu\nu}F_{\alpha\beta}\}{\cal D}\bar{\psi}{\cal D}\psi.
\end{eqnarray}
The anomaly is integrated for a finite transformation, and it 
gives rise to the so-called Wess-Zumino term~\cite{wess}.

Based on this observation, one recognizes the following 
differences between the geometric phases and chiral 
anomaly:~\cite{fujikawa5}\\
1. The Wess-Zumino term is added to the classical action in 
path integral, whereas the geometric term appears {\em inside} 
the classical action sandwiched by field variables as in (3). 
Geometric phases are thus state-dependent.
\\
2. The topology of chiral anomaly, which is provided by gauge 
fields, is exact, whereas the topology of the adiabatic 
geometric phase, which is valid only in the adiabatic limit, is 
trivial.\\

\end{document}